\documentclass[12pt]{article}
\usepackage{setspace}
\usepackage{indentfirst}
\usepackage{natbib}
\usepackage{xltabular}
\usepackage{booktabs}
\usepackage{array}
\usepackage{subcaption}
\newcolumntype{L}[1]{>{\raggedright\arraybackslash}p{#1}}
\newcolumntype{R}[1]{>{\raggedleft\arraybackslash}p{#1}}
\usepackage{amsmath}
\usepackage{ragged2e}
\usepackage{authblk}
\usepackage{amsmath}
\usepackage{breqn}
\usepackage{graphics}
\usepackage{amsfonts}
\usepackage{graphicx}
\usepackage{amsmath}
\usepackage{wrapfig}
\usepackage{booktabs,caption}
\usepackage{adjustbox}
\usepackage{enumitem}
\usepackage[flushleft]{threeparttable}
\usepackage[colorlinks,
            linkcolor=blue,
            anchorcolor=blue,
            citecolor=blue]{hyperref}
            
\usepackage[left=1.in, right=1.in, top=.5in, bottom=.5in, includehead, includefoot]{geometry}       

\usepackage{floatrow}
\usepackage[multiple]{footmisc}
\usepackage{caption}
\usepackage{rotating}

\begin{document}
	
\begin{titlepage}
\title{How Innovation Shapes Financial Structure: The Moderating Role of Institutional Quality\thanks{We would like to thank seminar participants at Waseda University and acknowledge valuable comments from Tuo Chen, Shenzhe Jiang, Munechika Katayama, Junko Koeda, Qing Liu, Similan Rujiwattanapong, Yueting Tong, Kozo Ueda, and Longtian Zhang. This study was supported by the Yu-cho Foundation (Grant-in-Aid for Research, 2024). Corresponding author: Tomoo Kikuchi. Nishi-Waseda Bldg.7F, 1-21-1 Nishi-Waseda, Shinjyuku-ku, Tokyo 169-0051 Japan. Email: \href{mailto:tomookikuchi@waseda.jp}{tomookikuchi@waseda.jp}}} 
\author[a]{Yimin Wu}
\author[b]{Tomoo Kikuchi}
\affil[a,b]{Graduate School of Asia-Pacific Studies, Waseda University}
\date{\today}
\maketitle

\begin{abstract}
\noindent This paper studies how the stock market---relative to the banking sector---responds to innovation  by using a panel of 75 countries from 1982 to 2021. Our baseline result is that innovation has a positive effect on stock market activity, efficiency and size relative to the banking sector. In addition, we uncover alternative funding channels by studying how institutional quality moderates the effect of innovation on financial structure. While the moderating effect is positive for activity and efficiency, it is negative for size, suggesting a larger role of banks under high institutional quality.  Furthermore, the moderating effect can be nonlinear for efficiency, suggesting alternative efficient funding channels under low institutional quality. 

\noindent\textbf{Keywords:} innovation; financial structure; institutional quality

\end{abstract}

\setcounter{page}{0}
\thispagestyle{empty}

\end{titlepage}

\pagebreak \newpage

\section{Introduction} \label{sec:introduction}

Most studies on the relationship between innovation and financial systems share two perspectives.
First, stock markets are better positioned to fund innovation than banks and second, innovation is a dependent variable, an outcome shaped by financial markets \citep{holmstrom1993market, allen1999diversity, hsu2014financial, brown2009financing, brown2013law, rajan1998finance, BoltonFreixas2000, kortum2000venture}.  
We contribute to the literature in two ways. First, unlike most previous studies, we study innovation as a driver of change in financial structure. Second,  we study how institutional quality moderates how innovation shapes financial structure in three dimensions---stock market activity, efficiency and size relative to the banking sector.

\begin{figure}[ht!]
	\centering
	\begin{subfigure}{0.49\textwidth}
		\includegraphics[width=\textwidth]{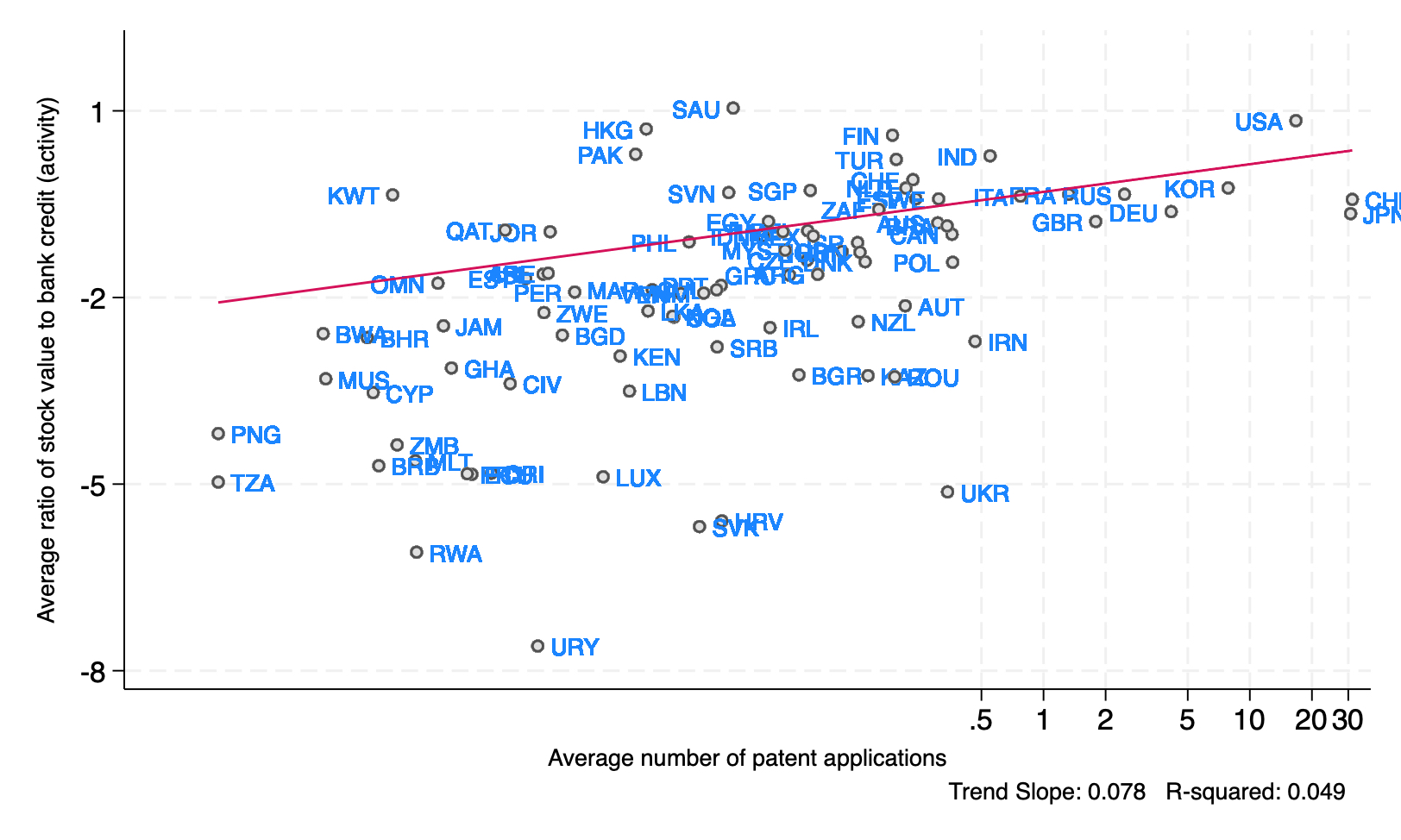}
		\caption{The activity dimension }
		\label{fig:activity}
	\end{subfigure}
	\hfill
	\begin{subfigure}{0.49\textwidth}
		\includegraphics[width=\textwidth]{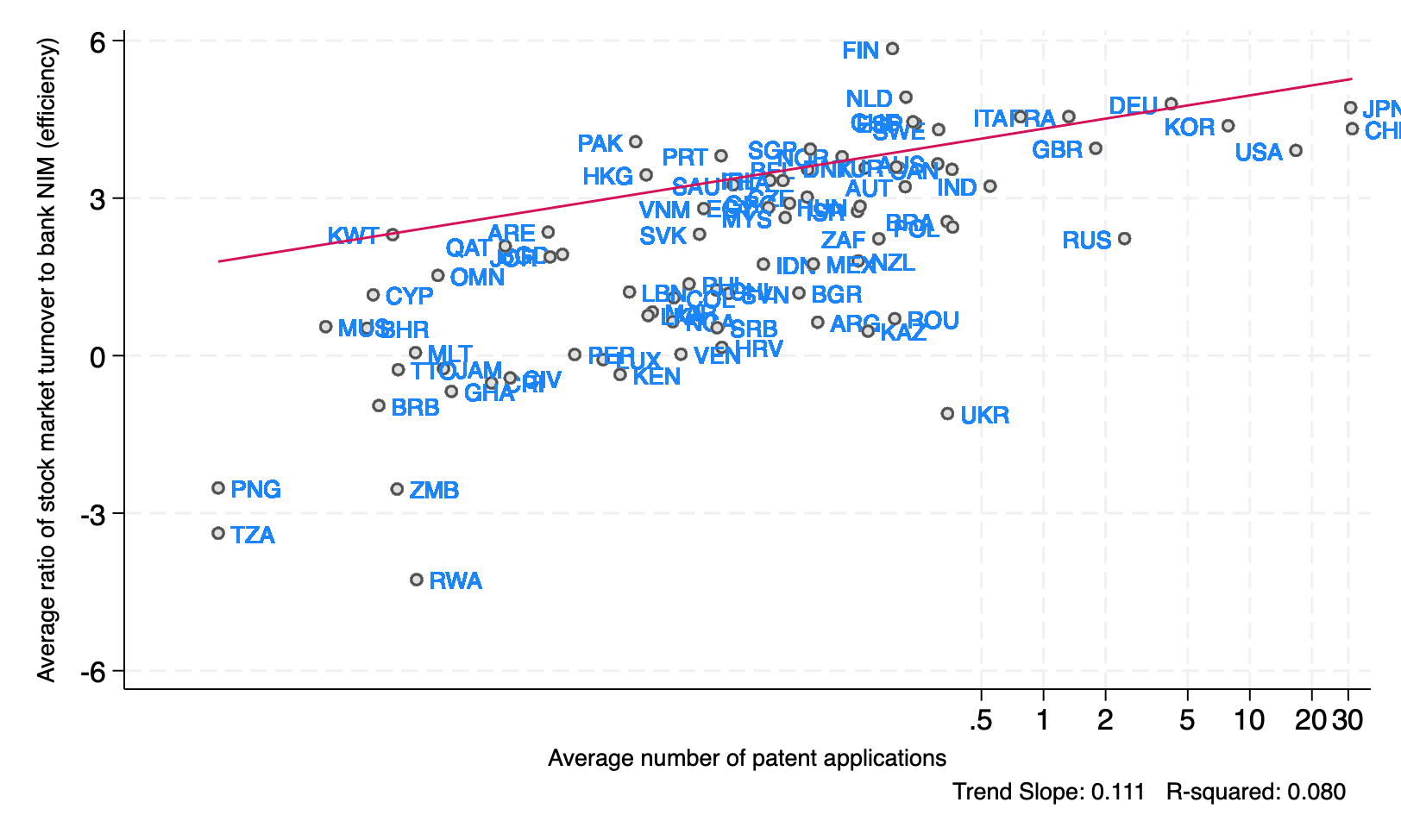}
		\caption{The efficiency dimension}
		\label{fig:efficiency}
	\end{subfigure}	
	\label{fig:activity_efficiency}
		\begin{subfigure}{0.49\textwidth}
		\includegraphics[width=\textwidth]{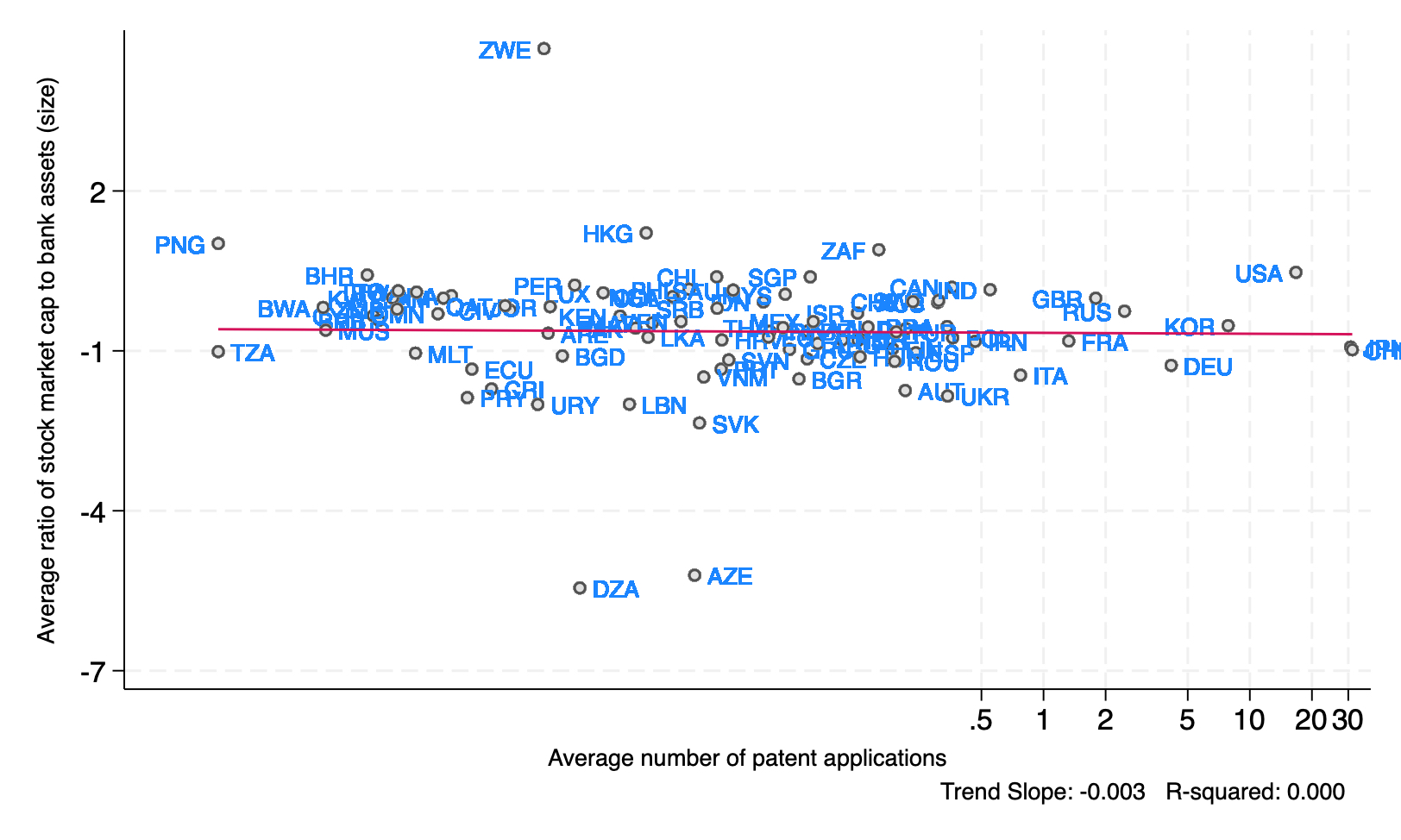}
		\caption{The size dimension}
		\label{fig:size}
	\end{subfigure}	
	\caption{Plot of innovation against three dimensions of financial structure}
	\label{fig:financial_structure}
\end{figure}

Figure \ref{fig:financial_structure} shows scatter plots---country averages from 1960 to 2021 for 159 countries---of innovation proxied by the number of patent applications against stock market activity, efficiency and size relative to the banking sector as defined in \citep{beck1999new, levine2002bank}. Higher ratios mean that stock markets are more active, more efficient or larger in size than the banking sector. 
We can see that stock markets are indeed more active and efficient than the banking sector in countries where innovation is higher, even though there is a significant variation at low level of innovation. On the other hand, there is almost no relationship between innovation and  stock market size relative to the banking sector with little variation in the relative size over innovation. 
Those stylized facts suggest that innovation might be positively associated with stock market activity and efficiency relative to the banking sector, but the association might not hold for stock market size relative to the banking sector. We will study the moderating role of institutional quality to further investigate the relationship between innovation and financial structure that can not be captured by simple two-dimensional scatter plots. 

The literature  traditionally views financial development as a factor mitigating financial constraints for entrepreneurs but not as a determinant causing innovation \citep{levine1997financial}.
For example, stock markets help companies grow and raise innovative capacity by financing R\&D investments \citep{brown2009financing}, but this requires innovative activity in the first place.   
On the theoretical front, it was shown that capital accumulation may expand equity financing by reducing monitoring \citep{boyd1998evolution} and bankruptcy costs \citep{bose2005endogenous}. 
Similarly, in our view, patent activity should make innovation outputs observable and verifiable and reduce information asymmetry, thereby compelling revaluation of firms in capital markets. Hence, surges in patent activity could be viewed as demand shocks for external finance that should raise stock market importance relative to the banking sector.

We use a panel of 75 advanced and emerging economies over 1982–2021 to show that innovation increases stock market activity, efficiency and size  relative to the banking sector and that those marginal effects are moderated by institutional quality, incorporating views from the law-finance nexus literature \citep{porta1998law, la2000investor, levine2000financial, brown2013law}. The demand shocks depend critically on institutional quality as markets may fail to process innovation signals with weak investor protection, poor enforcement or limited disclosure \citep{la2000investor}, whereas strong institutions enhance the ability of investors to interpret market signals and enforce contracts \citep{porta1998law, levine2000financial}.

Our main findings are as follows. First, the number of patent applications is positively associated with stock market activity, efficiency and size relative to the banking sector.  
To establish causality from innovation to financial structure, we interact a country's geographical distance to the regional innovation leader with regional patent growth and define the interaction terms as an IV, based on the idea that technology spills over regionally, while financial structure is determined by country-specific factors \citep{levine2023legal}.

Second, the marginal effect of innovation on financial structure is moderated by institutional quality indicators based on \cite{kaufmann2024worldwide}.\footnote{The institutional quality indicators include regulatory quality, voice and accountability, rule of law, political stability, corruption control, and government effectiveness.} In particular, we find that the moderating effects of institutional quality are heterogeneous. While institutional quality indicators, when significant, have a positive moderating effect on  stock market activity and efficiency relative to the banking sector, they have, when significant, a negative moderating effect on the relative size. 
This hints at the increasing  role of the banking sector  for financing innovation in economies with high institutional quality.

Third, the moderating effect of institutional quality can be nonlinear  for stock market efficiency relative to the banking sector. For efficiency, some institutional quality indicators (control of corruption, rule of law, and government effectiveness) exhibit a U-shaped moderating effect: innovation raises stock market efficiency relative to the banking sector at both low and high levels of institutional quality.
Since the banking sector's response to innovation  does not vary much with institutional quality, the stock market is mostly responsible for the nonlinear moderating effect. The result of a large moderating effect  at a low level of institutional quality suggests that there might  be alternative financing channels not captured by the literature on law-finance nexus but are relevant for economies with low institutional quality. One example of such a channel is based on reputation and relationships explored for China in \cite{allen2005law}.

Lastly, we use alternative innovation proxies, financial structure indicators and instruments 
to ensure that our findings are not driven by any single proxy and offer more robust and generalizable evidence on the role of innovation in shaping financial structure.

Regarding the functional differences between banks and stock markets in
financing innovation, our results are in line with theoretical arguments that banks are generally reluctant to finance high-risk, intangible R\&D activities \citep{holmstrom1993market, allen1999diversity}, and consistent with empirical findings that
stock markets and venture capital are better equipped than credit market 
to support innovation by pooling risk and valuing future growth potential \citep{rajan1998finance,kortum2000venture, brown2009financing, brown2013law, hsu2014financial}. However, as discussed above, we find that institutional quality moderates the relationship between innovation and financial in a way that the law-finance nexus literature does not capture.

The rest of the paper is organized as follows: Section \ref{data} describes the data. Section \ref{iv} presents the main results. 
Section \ref{sec:moderation} studies the moderating effect of institutional quality. Section \ref{sec:nonlinear} finds that the  moderating effect of institutional quality can be nonlinear. 
Section \ref{sec:conclusion} concludes.

\section{Data}\label{data}

We examine the impact of innovation on financial structure using an unbalanced panel of 159 advanced and emerging market economies, with annual data from 1960 to 2021. After including the necessary instruments and control variables, the estimated sample is reduced to 75 countries spanning from 1982 to 2021. We measure innovation of each country by the number of patent applications and stock market activity, efficiency and size relative to the banking sector. 
The definition and sources of the variables are given in Table \ref{table:data_description}. 

\begin{table}[ht!]
\begin{spacing}{0.5}
\centering
\footnotesize
\renewcommand{\arraystretch}{1} 
\begin{threeparttable}
\begin{tabular}{p{4cm} p{6.5cm} p{4cm}} 
\toprule
Notation & Description & Data Source \\
\midrule
\textbf{Dependent variables} & & \\
\hspace{0.2cm}$y^{act}_{i,t}$ & The ratio of stock value traded to the domestic credit to the private
	sector by banks.&  \cite{beck1999new,beck2009financial} \newline \cite{vcihak2012benchmarking} \\
\hspace{0.2cm}$y^{size}_{i,t}$ & The ratio of stock market capitalization to deposit money banks’ assets. &  \\
\hspace{0.2cm}$y^{eff}_{i,t}$ & The ratio of the stock market turnover ratio to the banking sector’s net interest margin. & \\
\midrule
\textbf{Independent variables} & & \\
\hspace{0.2cm}$x_{i,t}$ & The number of patent applications (10 thousands)  filed through the Patent Cooperation Treaty or national offices. & World Bank Database \\
\hspace{0.2cm}$q_{i,t}$ & One of six institutional quality measures: regulatory quality; voice and accountability; control of corruption; rule of law; political stability; or government effectiveness. & \cite{kaufmann2024worldwide} \\
\midrule
\textbf{Instruments} & & \\
\hspace{0.2cm}$z_{i,t}$ & The interaction term between the logarithm of the geographical distance between country $i$ and regional technological leader and average regional patent growth excluding country $i$. & GeoDist Database \newline World Bank Database \\
\hspace{0.2cm}$D_{i,r}$ & The set of regional dummies that equals 1 if the country $i$ belongs to region $r$, and 0 otherwise. Regions include Asia, Africa, and the Western Hemisphere. & \cite{liu2023capital} \\
\midrule
\textbf{Control variables} & & \\
\hspace{0.2cm}$Finopen_{i,t}$ & Index measuring capital account openness. & \cite{chinn2008new} \\
\hspace{0.2cm}$Tradopen_{i,t}$ & Sum of exports and imports of goods and services measured as a share of GDP. & OECD National Accounts \& World Bank Database \\
\hspace{0.2cm}$HCI_{i,t}$ & Years of schooling and returns to education. & Penn World Tables 10.01 \\
\hspace{0.2cm}$Goversize_{i,t}$ & Government final consumption expenditure excluding capital formation in defense and security. & World Bank Database \\
\hspace{0.2cm}$Inflat_{i,t}$ & Inflation measured by annual percentage change in consumer price index (Laspeyres formula). & World Bank Database \\
\hspace{0.2cm}$Gdpgrow_{i,t}$ & Real GDP growth rate. & IMF World Economic \newline Outlook Database \\
\hspace{0.2cm}$Gdpcap_{i,t}$ & GDP divided by midyear population, data are in constant 2015 U.S. dollars. & World Bank Database \\
\hspace{0.2cm}$Bankcri_{i,t}$ & Dummy indicating banking crisis: 1 if crisis, 0 otherwise. Defined by significant financial distress and policy intervention. & \cite{laeven2018systemic} \\
\bottomrule
\end{tabular}
\caption{Definition and notation of variables}
\label{table:data_description}
\end{threeparttable}
\end{spacing}

\end{table}

Table \ref{table:sum} reports the summary statistics for the full sample and for advanced and emerging economies separately. The average number of patent applications  per year is 2.21, with a high standard deviation of 8.59 and a maximum of 124.57. This indicates substantial cross-country heterogeneity, with innovation concentrated in a few technologically advanced economies. The distribution of patents is more clustered for advanced economies (standard deviation of 8.23) than emerging markets (8.77). In particular, the maximum number of patents in emerging economies (124.57) far exceeds that in advanced countries (38.74) despite a much smaller mean of 1.14 compare to 3.41 in advanced economies, driven by rapidly industrializing countries such as China.

Regarding financial structure, Table \ref{table:sum} shows that 
advanced economies have a higher mean of stock market {\em activity} and {\em efficiency} relative to the banking sector than emerging economies.  In contrast, emerging economies exhibit a higher mean of  stock market {\em size} relative to the banking sector than advanced economies, indicating that higher innovation activity is not necessarily associated with a larger stock market relative to the banking sector.  

 \begin{table}[ht!]
        \centering
        \footnotesize
            \begin{threeparttable}
                \begin{tabular}{p{5.3cm} R{1.5cm} R{1.5cm} R{1.5cm} R{1.5cm} R{1.5cm}} 
                    \toprule
                    Variable & N & Mean & S.D. & Min & Max \\ [0.5ex] 
                    \midrule
                     \textbf{Full sample} &  &  &  &  &  \\
                    \hspace{0.2cm}Patents & 1399 & 2.207 & 8.589 & 0.000 & 124.571 \\
                    \hspace{0.2cm}Financial structure (activity) & 1399 & 0.506 & 0.874 & 0.000 & 12.039 \\
                    \hspace{0.2cm}Financial structure (efficiency)& 824 & 32.682 & 49.391 & 0.033 & 436.470 \\
                      \hspace{0.2cm}Financial structure (size)& 1377 & 0.780 & 0.736 & 0.001 & 8.297 \\
                    \hspace{0.2cm}Financial openness & 1399 & 0.872 & 1.465 & -1.927 & 2.311 \\
                    \hspace{0.2cm}GDP growth & 1399 & 3.397 & 3.487 & -14.100 & 14.500 \\
                     \hspace{0.2cm}GDP per capita & 1399 & 19783.840 & 18462.370 & 755.482 & 84611.100 \\
                    \hspace{0.2cm}Trade openness & 1399 & 81.736 & 67.275 & 14.391 & 442.620 \\
                    \hspace{0.2cm}Inflation & 1399 & 16.170 & 124.053 & -4.009 & 2947.733 \\
                    \hspace{0.2cm}Government spend & 1399 & 16.089 & 4.720 & 4.403 & 29.322 \\
                    \hspace{0.2cm}Human capital & 1399 & 2.813 & 0.552 & 1.387 & 3.974 \\
                     \hspace{0.2cm}Bank crisis dummy & 1399 & 0.110 & 0.313 & 0 & 1 \\
                     \hspace{0.2cm}Regulatory quality & 1002 & 0.615 & 0.812 & -1.709 & 2.255 \\
                    \hspace{0.2cm}Voice and Accountability & 1005 & 0.388 & 0.909 & -1.907 & 1.801 \\
                    \hspace{0.2cm}Corruption Control& 1005 & 0.468 & 1.025 & -1.597 & 2.410 \\
                    \hspace{0.2cm}Rule of Law & 1005 & 0.506 & 0.923 & -1.202 & 2.026 \\
                    \hspace{0.2cm}Political Stability & 1005 & 0.086 & 0.965 & -2.810 & 1.753 \\
                    \hspace{0.2cm}Government Effectiveness & 1005 & 0.628 & 0.866 & -1.114 & 2.426 \\
                     \midrule
                    \textbf{Sub sample} &  &  &  &  &  \\
                    \hspace{0.2cm}Advanced economies &  &  &  &  &  \\
                    \hspace{0.5cm}Patents & 657 & 3.411 & 8.228 & 0.000 & 38.736 \\
                    \hspace{0.5cm}Financial structure (activity) & 657 & 0.594 & 0.920 & 0.000 & 12.039 \\
                    \hspace{0.5cm}Financial structure (efficiency) & 421 & 51.572 & 60.742 & 0.057 & 436.470 \\
                    \hspace{0.5cm}Financial structure (size)  & 788 & 0.741 & 0.774 & 0.022 & 8.297 \\
                    \hspace{0.2cm}Emerging economies &  &  &  &  &  \\
                    \hspace{0.5cm}Patents & 742 & 1.141 & 8.766 & 0.000 & 124.571 \\
                    \hspace{0.5cm}Financial structure (activity) & 742 & 0.434 & 0.828 & 0.000 & 11.038 \\
                    \hspace{0.5cm}Financial structure (efficiency) & 403 & 12.948 & 19.370 & 0.033 & 172.869 \\
                    \hspace{0.5cm}Financial structure (size) & 589 & 0.833 & 0.679 & 0.001 & 4.406 \\
                    \bottomrule
                \end{tabular}
                \begin{tablenotes}
                    \small
                    \item Note: Summary statistics of the data sample for the baseline regressions.
                \end{tablenotes}
                \caption{Summary Statistics}
                \label{table:sum}
            \end{threeparttable}
    \end{table}
 
\section{The panel instrumental variable approach}\label{iv}

This section presents our main empirical results. To support our claim for the direction of causality from innovation to financial structure, 
we construct an instrumental variable (IV) based on the idea that innovation diffusion is region-specific due to geographical proximity and trade links \citep{jaffe1993geographic, audretsch1996r}, and that regional characteristics significantly influence technological spillovers \citep{keller2002geographic, moretti2004workers, bloom2013identifying}.

\subsection{The baseline model}

Let $S_{r}$ denote the set of all countries in the region $r$. We separate the world into four regions: Africa, Asia, Europe and the Western Hemisphere. We first identify the regional innovation leader $l_r$ in each region $r$ as the country $j$ with the highest number of patent applications in a chosen base year $t_{1980}$, formally defined as
\begin{equation*}\label{eq:leader}
l_{r} = \arg \max_{j \in S_{r}} \ \text{pat}_{j, t_{1980}}.
\end{equation*}
The regional leaders based on equation above are South Africa (Africa), Japan (Asia), Germany (Europe) and the United States (Western Hemisphere). For each country $i$ in region $r$, we compute the average patent growth among all other countries in the same region except for country $i$, defined as
\begin{equation*}
g_{r, t}^{-i} 
=
\frac{1}{|S_{r}| - 1}
\sum_{\substack{j \in S_{r}, j \neq i}}
\Delta\text{pat}_{j, t}.
\end{equation*}
We then define $d_{i,l_r}$ as the logged geographical distance between the largest city of country $i$ and the regional leader $l_r$. These distances are calculated using the great circle formula, based on the latitudes and longitudes of the most important cities in each country \citep{head2002illusory}. Finally, the instrument for country $i$ in year $t$ is constructed as
\begin{equation*}
z_{i, t} = d_{i,l_r} \times g_{r, t}^{-i}.
\end{equation*}

The instrument is an exposure (shift-share) interaction between a predetermined exposure term $d_{i,l_r}$ and a leave-one-out shift $g_{r, t}^{-i}$.\footnote{Shift-share instruments can be justified under alternative exposure-based or shock-based identification frameworks, each of which requires additional orthogonality conditions \citep{goldsmith2020bartik, borusyak2022quasi}.} In our setting, the exposure term is the predetermined geographic distance between country $i$ and its regional innovation leader. Because it is fixed before the estimation period, it is not affected by contemporaneous domestic innovation or financial conditions. Likewise, the shift removes any mechanical contribution of domestic patenting to the regional shock. Therefore, we claim that, conditional on time fixed effects, regional dummies, and the observed controls, the interaction between distance and regional patent growth affects the composition of domestic finance only through its effect on domestic patenting.

We employ a 2SLS estimator to estimate the causal impact of innovation on financial structure. In the first stage, we regress current patent applications on our constructed instrument, regional dummies, control variables and time-fixed effects. The fitted values obtained represent the exogenous variation in innovation activity. In the second stage, we regress our measures of financial structure on these fitted innovation values, excluding country fixed effects since regional fixed effects are included as instruments \citep{liu2023capital}. This panel IV regression provides a structural interpretation that identifies the causal relationship between innovation and financial structure. The weak-instrument diagnostics reported confirm the validity and strength of our IV strategy. Furthermore, the coefficient estimates on innovation remain robust and statistically significant across specifications. This consistency suggests that our strategy effectively captures the exogenous variation, thereby supporting the conclusions drawn from our causal inferences in the empirical analysis.

Our baseline second-stage specification is given by
\begin{equation*}\label{eq:1}
	 \ln(y_{i,t})= \beta_1 x_{i,t}+ \beta\mathbf{W}_{i,t} + \mu_t + \varepsilon_{i,t}
\end{equation*}
where $y_{i,t}$ denotes stock market activity, efficiency or size  relative to the banking sector for country $i$ in year $t$, as defined in Table \ref{table:data_description} and $x_{i,t}$ is the number of patent applications for country $i$ in year $t$. We introduce a vector of conditioning variables denoted by $\mathbf{W}_{i,t}$ to control for other characteristics that may influence financial structure including years of schooling and returns to education, a banking crisis dummy, a measure of financial openness, the general government consumption, the annual change of the consumer price index (CPI), the real GDP growth rate, GDP per capita and the sum of exports and imports of goods and services. Given that regional innovation growth may correlate with global factors affecting domestic financial markets, we also control for time fixed effects $\mu_t$. $\varepsilon_{i,t}$ represents the regression residual, with standard errors clustered by year.

The first stage regression equation is given by 
\begin{equation*}
    x_{i,t} = \gamma_1 z_{i, t} + \sum_{r} \phi_r D_{r,i} + \gamma\mathbf{W}_{i,t} + \theta_t + \epsilon_{i,t}
\end{equation*}
where \( r \) indexes the regions (Asia, Africa or Western Hemisphere), \( \phi_r \) represents the coefficient for each region, \({D}_{r,i} \) is a dummy variable that equals 1 if country \( i \) belongs to region \( r \), and 0 otherwise,\footnote{Note that if a country belongs to Europe, all \({D}_{r,i} \) are 0.} and $z_{i, t}$ serves as an instrument that proxies regional innovation spillovers that affects the country $i$ in year $t$.

Table \ref{table:base} reports the results from our baseline IV estimations. The first-stage regressions confirm the strength of our instruments, as indicated by significant negative coefficients on the constructed instrument across specifications and first-stage F-statistics above conventional thresholds for instrument relevance with and without controls. Moreover, our weak-instrument test jointly helps to confirm the relevance of the instrument, as the p-values are consistently close to zero. In the second stage, the estimated coefficients on patent applications are positive and statistically significant across all three dimensions of financial structure.

\begin{table}[ht]
    \centering
    \footnotesize
    \begin{threeparttable}
    \begin{tabular}{
        >{\raggedright\arraybackslash}p{2.3cm} 
        >{\centering\arraybackslash}m{1.5cm} 
        >{\centering\arraybackslash}m{1.5cm}
        >{\centering\arraybackslash}m{1.5cm} 
        >{\centering\arraybackslash}m{1.5cm} 
        >{\centering\arraybackslash}m{1.5cm}
        >{\centering\arraybackslash}m{1.5cm}
    } 
         \toprule
         & \multicolumn{2}{c}{Activity} & \multicolumn{2}{c}{Efficiency} & \multicolumn{2}{c}{Size} \\
         \cmidrule(lr){2-3} \cmidrule(lr){4-5} \cmidrule(lr){6-7}
         & (1) & (2) & (3) & (4) & (5) & (6) \\
         \midrule
    
        \underline{2nd Stage} &  &  &  &  &  & \\
         
        Patents & 0.133**  & 0.178***  & 0.521***  & 0.699*** & 0.165** & 0.366*** \\
                & (0.057) & (0.062)  & (0.106) & (0.167) & (0.064) & (0.093)  \\
                
        \underline{1st Stage} &  &  &  &  &  & \\
        
        IV & -1.397***  & -0.956**  & -0.434*** & -0.208* & -0.278***  & -0.147**  \\
           & (0.263) & (0.381) & (0.134) & (0.114) & (0.091) & (0.065)  \\
           
        \underline{Weak IV Test} &  &  &  &  &  & \\ 
        
        CLR  & 
        0.000 & 0.000 &
        0.002 & 0.000 &
        0.000 & 0.000  \\
        
        AR  & 
        0.000 & 0.000 & 
        0.002 & 0.000 &
        0.000 & 0.000  \\
        
        Wald  & 
        0.019 & 0.004 & 
        0.000 & 0.000 & 
        0.010 & 0.000 \\
        
        1st F-statistic & 
        26.483 & 18.427 & 
        19.290 & 18.695 & 
        20.057 & 18.636  \\
        
        \midrule
        Period & 1982--2021 & 1982--2021 & 1991--2021 & 1991--2021 & 1982--2021 & 1982--2021 \\
        Time FE & YES & YES & YES & YES & YES & YES \\
        Controls & YES & NO & YES & NO & YES & NO \\
        Obs. & 1,123 & 1,499 & 693 & 959 & 1,114 & 1,479 \\
       [1ex] 
       \bottomrule
    \end{tabular}
    \begin{tablenotes}
        \footnotesize
        \item The dependent variable is one of three measures for financial structure. The endogenous variable is the number of patent applications instrumented with the regional innovation spillover instrument and regional dummies. Control variables are in Table \ref{table:data_description}. Note: *** \( p < 0.01 \), ** \( p < 0.05 \), * \( p < 0.1 \). Numbers in parentheses are standard errors clustered by year. AR and Wald tests follow the procedures in \cite{olea2013robust}. Multiple IVs yield extra CLR statistics; see \cite{pflueger2015robust} for discussions of weak instrument tests in linear IV regressions and \cite{finlay2014weakiv10} for Stata implementations. \(P\)-values are reported for CLR, AR, and Wald tests.
    \end{tablenotes}
    \caption{Effect of innovation on financial structure: Full sample}
    \label{table:base}

\end{threeparttable}
\end{table}

Specifically, Columns (1) and (2) show that the marginal effect on stock market activity  relative to the banking sector is 0.13 and 0.18, significant at the 5\% and 1\% levels with and without controls.\footnote{This means additional 10,000 patent applications lead on average to a 14\%-20\% increase in a country's ratio of stock value traded to bank credit. When $y=e^{\beta x_t}$, $\frac{y_{t+1}-y_t}{y_t}=\frac{e^{\beta x_{t+1}}-e^{\beta x_t}}{e^{\beta x_t}}=e^{\beta(x_{t+1}-x_t)}-1$. Hence, a unit change in $x_t$, i.e., $x_{t+1}-x_t=1$, leads to  $\frac{y_{t+1}-y_t}{y_t}=e^{\beta}-1$.}
Columns (3) and (4) show that the marginal effect on efficiency is the strongest, with coefficients 0.52 and 0.70, with and without controls, significant at the 1\% level. 
Columns (5) and (6) show the marginal effect on size is 0.17 and 0.37, significant at the 5\% and 1\% levels with and without controls. 

The consistent significance and positive signs of the estimated coefficients for all dimensions of financial structure confirm the hypothesis that innovation leads to a more active, more efficient and larger stock market relative to the banking sector. Moreover, our CLR, AR, and Wald statistics continue to reject the null hypothesis of weak instruments at 5\% significance level.

\subsection{Alternative innovation proxies}

To avoid relying solely on patent applications as the measure of innovation, we further incorporate a broader set of innovation proxies at the country level in Table \ref{table:innov}. Specifically, we employ six alternative indicators: (1) research and development (R\&D) expenditure; (2) the number of patents (in thousands) from OECD database; (3) total charges (in billion USD) for the authorized use of intellectual property (IP) rights, including patents, trademarks, copyrights, trade secrets and industrial processes; (4) high-technology exports (in billion USD), encompassing products with high R\&D intensity, such as aerospace, computers, pharmaceuticals and scientific instruments; (5) the number of scientific and technical publications indexed in SCI and SSCI journals, calculated using fractional author attribution across countries; and (6) an extended measure of patent applications that includes both resident and non-resident filings. Across all these alternative measures, our findings remain robust and consistent with the baseline results using the same identification strategy, which supports the role of innovation in shaping financial structure.

Table \ref{table:innov} shows that alternative innovation measures, R\&D expenditure, patent grants, IP receipts, high-tech exports and scientific papers, consistently exhibit positive and significant effects on stock market activity, efficiency and size relative to the banking sector,  though magnitudes vary. R\&D expenditure demonstrates the strongest impacts in general. High-tech exports have smaller but significant effects, suggesting a weaker direct impact. In general, comprehensive innovation indicators capture that innovation robustly increases  stock market activity, efficiency and size relative to the banking sector. 

\begin{sidewaystable}	
	\centering
	\footnotesize
	\makebox[\textwidth][c]{%
		\begin{threeparttable}
			\begin{tabular}{
					p{2.1cm} 
					>{\centering\arraybackslash}m{0.5cm} 
					>{\centering\arraybackslash}m{0.5cm}
					>{\centering\arraybackslash}m{0.5cm}
					>{\centering\arraybackslash}m{0.5cm}
					>{\centering\arraybackslash}m{0.5cm}
					>{\centering\arraybackslash}m{0.5cm}
					>{\centering\arraybackslash}m{0.5cm}
					>{\centering\arraybackslash}m{0.5cm}
					>{\centering\arraybackslash}m{0.5cm}
					>{\centering\arraybackslash}m{0.5cm}
					>{\centering\arraybackslash}m{0.5cm}
					>{\centering\arraybackslash}m{0.5cm}
					>{\centering\arraybackslash}m{0.5cm}
					>{\centering\arraybackslash}m{0.5cm}
					>{\centering\arraybackslash}m{0.5cm}
					>{\centering\arraybackslash}m{0.5cm}
					>{\centering\arraybackslash}m{0.5cm}
					>{\centering\arraybackslash}m{0.5cm}
				} 
				\toprule
				& \multicolumn{3}{c}{R\&D} 
				& \multicolumn{3}{c}{Patent Grants} 
				& \multicolumn{3}{c}{IP Receipts} 
				& \multicolumn{3}{c}{High-tech Exports} 
				& \multicolumn{3}{c}{Scientific Papers} 
				& \multicolumn{3}{c}{Total Patents} \\
				\cmidrule(lr){2-4} \cmidrule(lr){5-7} \cmidrule(lr){8-10} \cmidrule(lr){11-13} \cmidrule(lr){14-16} \cmidrule(lr){17-19}
				& (1a) & (1b) & (1c) & (2a) & (2b) & (2c) & (3a) & (3b) & (3c) & (4a) & (4b)& (4c) & (5a) & (5b) & (5c) & (6a) & (6b) & (6c) \\
				\midrule
				\underline{2nd Stage} & & & & & & & & & & & & & & & & & & \\
				Activity & 
				1.780***  &  &  & 
				0.761***  &  &  &
				0.959***  &  &  & 
				0.007**  &  &  &
				0.800***  &  &  &
				0.116***  &  &  \\
				
				& 
				(0.136) &  &  & 
				(0.118) &  &  &
				(0.118) &  &  & 
				(0.003) &  &  &
				(0.074) &  &  &
				(0.041) &  &  \\
				
				Efficiency & 
				&   1.460*** & & 
				&   0.642*** & &
				&   0.925*** & & 
				&   0.014*** & &
				&   1.290*** & &
				&   0.246*** & \\
				
				& 
				&    (0.130) && 
				&    (0.111) &&
				&    (0.091) && 
				&    (0.005) &&
				&    (0.073) &&
				&    (0.044) & \\
				
				Size & 
				&& 0.392***  &   
				&& 0.231***  &  
				&& 0.221***  &   
				&& 0.005  &
				&& 0.115*  &
				&& 0.203***    \\
				
				& 
				&& (0.061)   & 
				&&  (0.088)   &
				&&  (0.054)   & 
				&&  (0.003)   &
				&&  (0.063)   &
				&&  (0.036)   \\
				
				\underline{Weak IV Test} & & & & & & & & & & & & & & & & & & \\
				CLR & 
				0.001 & 0.003 &0.008   & 
				0.000 & 0.002 &0.000   &
				0.000 & 0.002 &0.000   &
				0.051 & 0.047 &0.043   &
				0.001 & 0.002 &0.001   &
				0.000 & 0.002 &0.000   \\
				
				AR & 
				0.001 & 0.002 &0.004   & 
				0.000 & 0.002 &0.000   &
				0.000 & 0.002 &0.000   &
				0.046 & 0.044 &0.040   &
				0.001 & 0.002 &0.001   &
				0.000 & 0.002 &0.000   \\
				
				Wald & 
				0.000 & 0.000 & 0.000 & 
				0.000 & 0.000 & 0.009 &
				0.000 & 0.000 & 0.000 &
				0.024 & 0.003 & 0.133 &
				0.000 & 0.000 & 0.067 &
				0.004 & 0.000 & 0.000 \\
				\midrule
				Observations & 
				687 & 551 &671   & 
				1,071 & 657 &1,066   &
				843 & 576 &847   &
				429 & 384 &406   &
				890 & 668 &854   &
				1,123 & 693 &1,114   \\
				
				Time period & 
				1997-2021 & 1997-2021 &1997-2021   & 
				1982-2021 & 1991-2021 &1982-2021   &
				1982-2021 & 1997-2020 &1982-2021   & 
				2008-2021 & 2008-2021 &2008-2021   &
				1997-2021 & 1997-2021 &1997-2021   & 
				1982-2021 & 1991-2021 &1982-2021   \\
				
				Time FE & 
				YES & YES &
				YES & YES &
				YES & YES &
				YES & YES &
				YES & YES &
				YES & YES &
				YES & YES &
				YES & YES &
				YES & YES \\
				
				Controls & 
				YES & YES &
				YES & YES &
				YES & YES &
				YES & YES & 
				YES & YES &
				YES & YES &
				YES & YES &
				YES & YES &
				YES & YES \\[1ex] 
				\bottomrule
			\end{tabular}
			\begin{tablenotes}
				\footnotesize
				\item The dependent variable is one of three dimensions for financial structure. The endogenous variable is the alternative innovation proxy instrumented with the regional innovation spillover instrument and regional dummies. Columns (1a) to (1c) report research and development (R\&D) expenditure. Columns (2a) to (2c) report the number of patents granted (in thousands) from the OECD database. Columns (3a) to (3c) show receipts for intellectual property rights in billion USD. Columns (4a) to (4c) report high-tech exports in billion USD. Columns (5a) to (5c) report counts of scientific and technical articles. Columns (6a) to (6c) show total patent applications for both residents and non-residents. Statistical significance: *** \( p < 0.01 \), ** \( p < 0.05 \), * \( p < 0.1 \). Standard errors clustered by year in parentheses. AR and Wald tests follow \cite{olea2013robust}. See \cite{pflueger2015robust} for discussions of weak instrument tests and \cite{finlay2014weakiv10} for Stata implementations. $P$-values are reported for CLR, AR, and Wald tests of weak instruments.
			\end{tablenotes}
			\caption{Alternative measures of innovation}
			\label{table:innov}
	\end{threeparttable}}
\end{sidewaystable}
\clearpage

\section{Moderating effect of institutional quality\label{sec:moderation}}

The preceding analysis demonstrates that the effect of innovation on  stock market activity, efficiency and size relative to the banking sector is robust. However, countries at similar levels of innovation capability may exhibit different outcomes if institutional quality differs. Institutions influence how effectively innovation is translated into market activity. Strong institutions promote legal certainty, regulatory quality and transparency, all of which are crucial for fostering investor confidence and enabling financial markets to respond to innovative activities.

Therefore, this section extends our empirical framework to examine whether institutional quality moderates the marginal effect of innovation on financial structure. We analyze six key institutional indicators: regulatory quality, voice and accountability, corruption control, rule of law, political stability and government effectiveness based on \cite{kaufmann2024worldwide}. Our second-stage empirical specification remains consistent with the approach used in the previous section such that
\begin{equation*}
\ln(y_{i,t}) = \sigma_1 x_{i,t} + \sigma_2 q_{i,t} + \sigma_3 x_{i,t} \times q_{i,t} + \sigma \mathbf{W}_{i,t} + \mu_t + \varepsilon_{i,t}
\end{equation*}
where $q_{i,t}$ represents one of the institutional quality measures for country $i$ in year $t$. 

Table \ref{table:inst_act} reports the results of the panel interaction model with instruments that examine how institutional quality moderates the marginal effect of innovation on stock market \emph{activity} relative to the banking sector. 
The coefficients of the interaction terms are positive and significant at the 1\% level for 
all institutional indicators. The weak-instrument tests (CLR, AR, and Wald) uniformly produce low p-values across all specifications and are significant at the 1\% level, supporting the strength of the instruments used.
However, the size of the coefficients varies. 
Notably, the interaction with voice and accountability presents the smallest coefficient of 0.391 (Column 2). On the other hand, the interaction with corruption control has the largest coefficient of 1.730 (Column 3).

\begin{table}[ht!]
	\centering
	\footnotesize
	\begin{threeparttable}
		\begin{tabular}{
				>{\raggedright\arraybackslash}p{3cm} 
				>{\centering\arraybackslash}m{1.8cm}
				>{\centering\arraybackslash}m{1.8cm}
				>{\centering\arraybackslash}m{1.8cm}
				>{\centering\arraybackslash}m{1.8cm}
				>{\centering\arraybackslash}m{1.8cm}
				>{\centering\arraybackslash}m{1.8cm}
			} 
			\toprule
			& \multicolumn{6}{c}{Activity} \\
			\cmidrule(lr){2-7}
			& (1) & (2) & (3) & (4) & (5) & (6) \\
			\midrule
			
			\underline{2nd Stage} &  &  &  &  & & \\
			Patents*Regulation & 
			1.227*** &  &  &  & & \\
			& (0.133)  &  &  & & \\
			
			Patents*Voice & 
			& 0.391***  &  &  & & \\
			&  & (0.031)  &  & & \\
			
			Patents*Corruption & 
			&  & 1.730***  &  &  &\\
			&  & & (0.233) &  & \\

			Patents*Law & 
			&  &  & 0.949***  &  &\\
			&  &  & & (0.110) &\\

			Patents*Politics & 
			&  &  &  & 1.314*** &\\
			&  &  &  & &(0.189) &\\
			
			Patents*Government & 
			&  &  &  &  &1.720*** \\
			&  &  &  & & & (0.325) \\
			
			\underline{Weak IV Test} &  &  &  &  & &\\
			
			CLR  & 
			0.004 & 0.004 & 0.004 & 0.005 & 0.005 & 0.005 \\
			
			AR  & 
			0.004 & 0.003 & 0.004 & 0.004 & 0.004 & 0.004 \\ 
			
			Wald  & 
			0.000 & 0.000 & 0.000 & 0.000 & 0.000 & 0.000 \\
			
			\midrule
			Period & 1998-2021 & 1998-2021 & 1998-2021 & 1998-2021 & 1998-2021  & 1998-2021 \\
			Time FE & YES & YES & YES & YES & YES & YES \\
			Controls & YES & YES & YES & YES & YES & YES \\
			Obs. & 805 & 808 & 808 & 808 & 808 & 808  \\[1ex] 
			\bottomrule
		\end{tabular}
		\begin{tablenotes}
			\footnotesize
			\item The dependent variable is the activity dimension of the financial structure. The endogenous variable is the number of patent applications instrumented with the regional innovation spillover instrument and regional dummies. Control variables are in Table \ref{table:data_description}. Note: *** \( p < 0.01 \), ** \( p < 0.05 \), * \( p < 0.1 \). Numbers in parentheses are standard errors clustered by year. AR and Wald tests follow the procedures in \cite{olea2013robust}. Multiple IVs yield extra CLR statistics; see \cite{pflueger2015robust} for discussions of weak instrument tests in linear IV regressions and \cite{finlay2014weakiv10} for Stata implementations. $P$-values are reported for CLR, AR, and Wald tests. F-statistics may have questionable accuracy in regressions with more than one endogenous regressor; thus, we rely on weak IV test results for our interaction regressions. 
		\end{tablenotes}
		\caption{Six Interacting institutional quality indicators  for financial structure (activity)}
		\label{table:inst_act}
	\end{threeparttable}
\end{table}

Table \ref{table:inst_eff} reports  how institutional quality moderates the marginal effect of innovation on stock market  \emph{efficiency} relative to the banking sector.
Specifically, the interaction with regulatory quality yields a coefficient of 0.982, with voice and accountability 0.330 and  with control of corruption 0.779, all significant at the 1\% level.   
The coefficients of the interaction terms are not significant for the rule of law, political stability and government effectiveness (Columns 4–6). The weak-instrument tests (CLR, AR, and Wald) uniformly produce low p-values across all specifications and are significant at the 5\% level, supporting the strength of the instruments used.

\begin{table}[ht!]
	\centering
	\footnotesize
	\begin{threeparttable}
		\begin{tabular}{
				>{\raggedright\arraybackslash}p{3cm} 
				>{\centering\arraybackslash}m{1.8cm}
				>{\centering\arraybackslash}m{1.8cm}
				>{\centering\arraybackslash}m{1.8cm}
				>{\centering\arraybackslash}m{1.8cm}
				>{\centering\arraybackslash}m{1.8cm}
				>{\centering\arraybackslash}m{1.8cm}
			} 
			\toprule
			& \multicolumn{6}{c}{Efficiency} \\
			\cmidrule(lr){2-7}
			& (1) & (2) & (3) & (4) & (5) & (6)\\
			\midrule
			
			\underline{2nd Stage} &  &  &  &  & &\\
			
			Patents*Regulation & 
			0.982***  &  &  &  &  &\\
			& (0.347) &  &  &  &\\
			
			Patents*Voice & 
			& 0.330***  &  &  &  &\\
			&  & (0.095) &  &  &\\
			
			Patents*Corruption & 
			&  &  0.779** &  &  &\\
			&  & & (0.334) &  &\\
			
			Patents*Law & 
			&  &  & 0.154  &  &\\
			&  &  &  & (0.267)  &\\
			
			Patents*Politics & 
			&  &  &  & -0.419 &\\
			&  &  &  & & (0.366) &\\
			Patents*Government & 
			&  &  &  & & -0.037 \\
			&  &  &  & &  & (0.291)\\
			
			\underline{Weak IV Test} &  &  &  &  & &\\
			
			CLR  & 
			0.008 & 0.007 & 0.008 & 0.009 & 0.008 &0.010\\ 
			
			AR  & 
			0.007 & 0.006 & 0.007 & 0.008 & 0.007 &0.009\\ 
			
			Wald  & 
			0.000 & 0.000 & 0.000 & 0.000 & 0.000 &0.000\\ 
			
			\midrule
			Period & 1998-2021 & 1998-2021 & 1998-2021 & 1998-2021 & 1998-2021 &1998-2021\\
			Time FE & YES & YES & YES & YES & YES &YES\\
			Controls & YES & YES & YES & YES & YES &YES\\
			Obs. & 645 & 649 & 649 & 649 & 649 &649 \\[1ex] 
			\bottomrule
		\end{tabular}
		\begin{tablenotes}
			\footnotesize
			\item The dependent variable is the efficiency dimension of financial structure. The endogenous variable is the number of patent applications instrumented with the regional innovation spillover instrument and regional dummies. Control variables are in Table \ref{table:data_description}. Note: *** \( p < 0.01 \), ** \( p < 0.05 \), * \( p < 0.1 \). Numbers in parentheses are standard errors clustered by year. AR and Wald tests follow the procedures in \cite{olea2013robust}. Multiple IVs yield extra CLR statistics; see \cite{pflueger2015robust} for discussions of weak instrument tests in linear IV regressions and \cite{finlay2014weakiv10} for Stata implementations. $P$-values are reported for CLR, AR, and Wald tests. F-statistics may have questionable accuracy in regressions with more than one endogenous regressor; thus, we rely on weak IV test results for our interaction regressions. 
		\end{tablenotes}
		\caption{Six interacting  institutional quality indicators  for financial structure (efficiency) }
		\label{table:inst_eff}
	\end{threeparttable}
\end{table}

Table \ref{table:inst_size} reports  how institutional quality moderates the marginal effect of innovation on stock market \emph{size} relative to the banking sector.  The interactions with regulatory quality and voice and accountability are negative but not significant (Columns 1-2). 
The interactions with control of corruption and government effectiveness demonstrate similar effects, with coefficients of -1.057 (Column 3) and -1.103 (Column 6), significant at the 1\% level. The interaction with political stability yields the largest magnitude of -1.367 (Column 5), significant at the 1\% level. Finally, the interaction with the rule of law yields the smallest among the significant interactions -0.527 (Column 4),  significant at the 1\% level.
The weak-instrument tests (CLR, AR) uniformly produce low p-values across all specifications and are significant at the 5\% level, supporting the strength of the instruments used.

\begin{table}[ht!]
	\centering
	\footnotesize
	\begin{threeparttable}
		\begin{tabular}{
				>{\raggedright\arraybackslash}p{3cm} 
				>{\centering\arraybackslash}m{1.8cm}
				>{\centering\arraybackslash}m{1.8cm}
				>{\centering\arraybackslash}m{1.8cm}
				>{\centering\arraybackslash}m{1.8cm}
				>{\centering\arraybackslash}m{1.8cm}
				>{\centering\arraybackslash}m{1.8cm}
			} 
			\toprule
			& \multicolumn{6}{c}{Size} \\
			\cmidrule(lr){2-7}
			& (1) & (2) & (3) & (4) & (5) & (6) \\
			\midrule
			
			\underline{2nd Stage} &  &  &  &  & & \\

			Patents*Regulation & 
			-0.191   &  &  &  &  &\\
			&(0.215)  &  &  &  &\\
			
			Patents*Voice & 
			& -0.066  &  &  &  &\\
			& & (0.058) &  &  &\\
			
			Patents*Corruption & 
			&  &  -1.057*** &  &  &\\
			&  &  & (0.219) &  &\\
			
			Patents*Law & 
			&  &  & -0.527***  &  &\\
			&  &  & & (0.189)  &\\
			
			Patents*Politics & 
			&  &  &  & -1.367***  &\\
			&  &  &  &  &(0.333)&\\
			
			Patents*Government & 
			&  &  &  & & -1.103*** \\
			&  &  &  &  & & (0.228)\\
			
			\underline{Weak IV Test} &  &  &  &  & &\\
			
			CLR  & 
			0.005 & 0.005 & 0.004 & 0.004 & 0.005 &0.004\\ 
			
			AR  & 
			0.004 & 0.004 & 0.004 & 0.004 & 0.004 &0.004\\ 
			
			Wald  & 
			0.060 & 0.176 & 0.000 & 0.019 & 0.000 &0.000\\ 
			
			\midrule
			Period & 1998-2021 & 1998-2021 & 1998-2021 & 1998-2021 & 1998-2021 &1998-2021\\
			Time FE & YES & YES & YES & YES & YES &YES\\
			Controls & YES & YES & YES & YES & YES &YES\\
			Obs. & 762 & 769 & 769 & 769 & 769 & 769\\[1ex] 
			\bottomrule
		\end{tabular}
		\begin{tablenotes}
			\footnotesize
			\item The dependent variable is the size dimension of financial structure. The endogenous variable is the number of patent applications instrumented with the regional innovation spillover instrument and regional dummies. Control variables are in Table \ref{table:data_description}. Note: *** \( p < 0.01 \), ** \( p < 0.05 \), * \( p < 0.1 \). Numbers in parentheses are standard errors clustered by year. AR and Wald tests follow the procedures in \cite{olea2013robust}. Multiple IVs yield extra CLR statistics; see \cite{pflueger2015robust} for discussions of weak instrument tests in linear IV regressions and \cite{finlay2014weakiv10} for Stata implementations. $P$-values are reported for CLR, AR, and Wald tests. F-statistics may have questionable accuracy in regressions with more than one endogenous regressor; thus, we rely on weak IV test results for our interaction regressions. 
		\end{tablenotes}
		\caption{Six interacting institutional quality indicators for financial structure (size) }
		\label{table:inst_size}
	\end{threeparttable}
\end{table}

Overall, the results in Table \ref{table:inst_act}-\ref{table:inst_eff} show that 
the moderating effect of 
institutional indicators is positive for activity and efficiency. 
This means that when the number of patent applications rises, stock market activity and  efficiency relative to the banking sector increase more in countries with higher institutional quality. This is consistent with evidence that investor protection and contract enforcement activate stock markets \citep{la1997legal}, and with the view that better governance lowers information and enforcement frictions, raising market liquidity \citep{demirgucc2001financial}.

In contrast, the results in Table  \ref{table:inst_size}  show that the moderating effect of 
institutional indicators is negative for the size dimension. In other words, 
institutional quality does not help innovation expand stock market size relative to the banking sector. This might reflect distinct roles of markets and banks in financing innovation. Stock markets are better suited than banks at funding new, uncertain projects, disseminating information, and facilitating risk-sharing for new technologies \citep{allen2000comparing}. This improves trading volume and market efficiency as investors speculate on and respond to patent-related news \citep{kogan2017technological}. In contrast, banks are better suited to scale up established innovations, providing large capital for expansion once new ideas mature. Thus, under strong institutions, banks actively finance innovative firms, often using patents as collateral for loans, which increases banks' assets \citep{robb2014capital,mann2018creditor}. Furthermore, many countries with high institutional quality often retain large banking sectors. Thus, stock market capitalization may not increase relative to banks' assets when innovation accelerates, because innovative firms often rely on bank credit and internal funds rather than issuing equity in high institutional quality economies \citep{cornaggia2015does,chava2017lending,spatareanu2019bank}.

\section{Nonlinear moderating effect \label{sec:nonlinear}}

The linear interaction $x_{i,t}\times{q}_{i,t}$ assumes that the marginal effect of innovation on financial structure varies linearly with institutional quality. 
To test for a possible nonlinear relationship, we augment the interaction with a quadratic term in institutional quality and its interaction with innovation (e.g., $x_{i,t}\times{q}_{i,t}^2$) following \cite{lind2010or}. To capture the nonlinear effects of institutional quality on the relationship between the number of patent applications and financial structure, we estimate 
\begin{equation*}
\ln(y_{i,t}) = \pi_1{x}_{i,t} + \pi_2({x}_{i,t}\times{q}_{i,t})
+ \pi_3({x}_{i,t}\times{q}_{i,t}^2) +\pi_4{q}_{i,t} + \pi_5{q}_{i,t}^2 + \pi\mathbf{W}_{i,t} + \mu_t + \varepsilon_{i,t}.
\end{equation*}
The marginal effect of innovation is
\begin{equation*}
    g({q}) \equiv \frac{\partial \ln(y_{i,t})}{\partial {x}_{i,t}}
= \pi_{1}+\pi_{2}{q_{it}}+\pi_{3}{q_{it}}^{2}.
\end{equation*}

If $\pi_{3}=0$, this slope varies linearly with institutional quality. If $\pi_{3}\neq 0$, the slope changes with institutional quality: $\pi_{3}>0$ can produce a U-shape in $g({q})$, while $\pi_{3}<0$ can produce an inverted-U. However, to confirm a presence of U (or inverted-U), we need: (i) $\pi_{3}$ statistically different from zero; (ii) the endpoint slopes $g'(L)$ and $g'(U)$ are significant and have opposite signs: $g'(L)<0<g'(U)$ for U shape and $g'(U)<0<g'(L)$ for inverted-U shape; and (iii)  the turning point ${q}^\star = -\frac{\pi_{2}}{2\pi_{3}}$ lies within $(L,U)$, where $L \equiv \min({q}_{i,t})$ and $U \equiv \max({q}_{i,t})$ are taken over the estimation sample (i.e., the observations used in the regression). 
Beyond establishing the existence of the (inverted) U-shaped relationship, we further consider the 95\% pointwise confidence band of the marginal effect to identify whether  the effect is statistically significant. We find that institutional quality has a robust nonlinear moderating effect (U-shaped) only for stock market efficiency.

We might be concerned that the nonlinear moderating effect is driven by a handful of exceptionally patent-intensive economies rather than by a systematic relationship.
This concern is most acute for the low-institution arm of the U-shape, since the influential evidence on reputation- and relationship-based financing under weak formal institutions is itself drawn
from China \citep{allen2005law} and India \citep{allen2012financing}, the two dominant patent filers among emerging markets in our sample. 
We therefore estimate the quadratic interaction specification on a sample that excludes China and India.

\begin{figure}[ht!]
	\centering
	\begin{subfigure}{0.49\textwidth}
		\includegraphics[width=\textwidth]{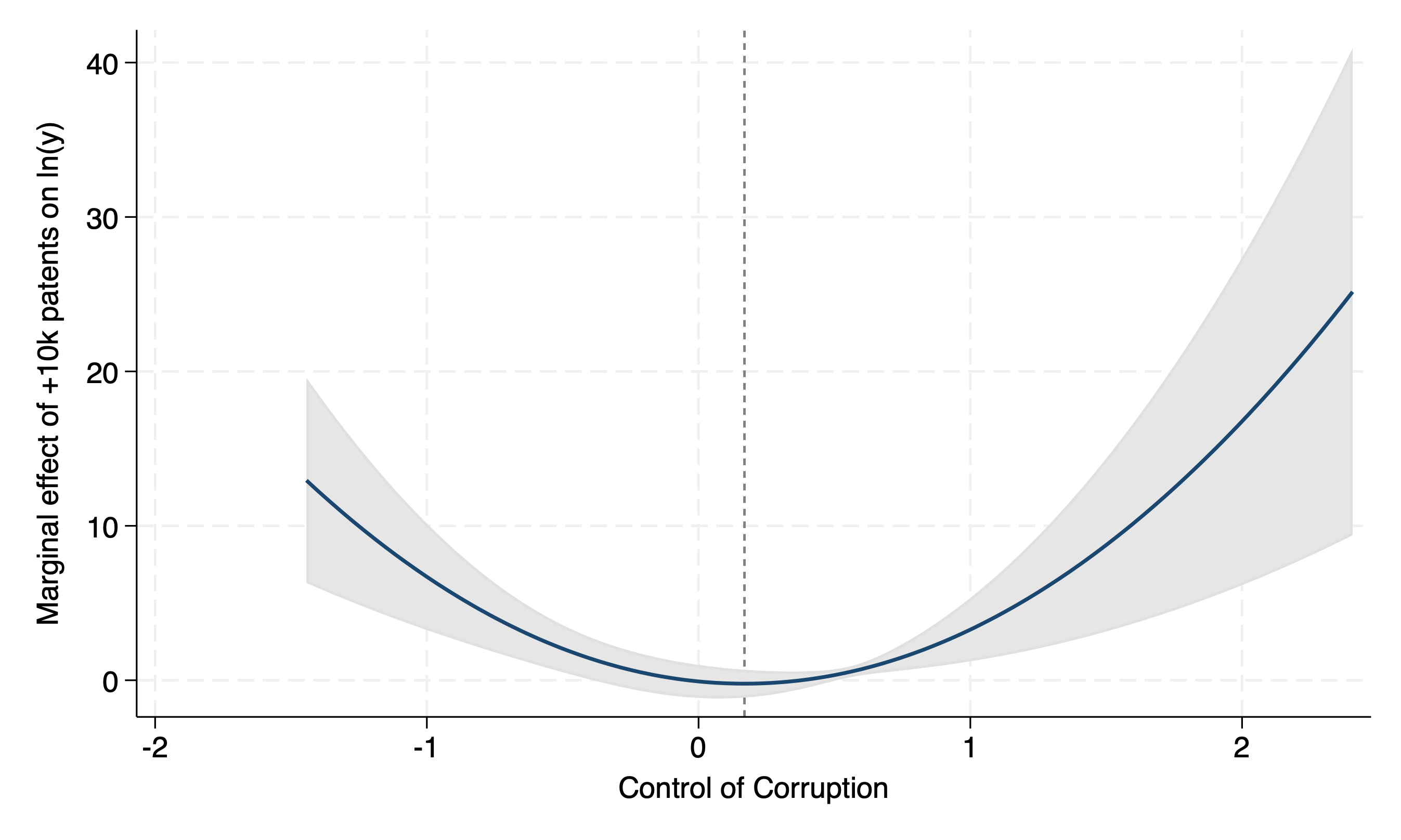}
		\caption{Control of corruption}
		\label{fig:effcor}
	\end{subfigure}
	\hfill
	\begin{subfigure}{0.49\textwidth}
		\includegraphics[width=\textwidth]{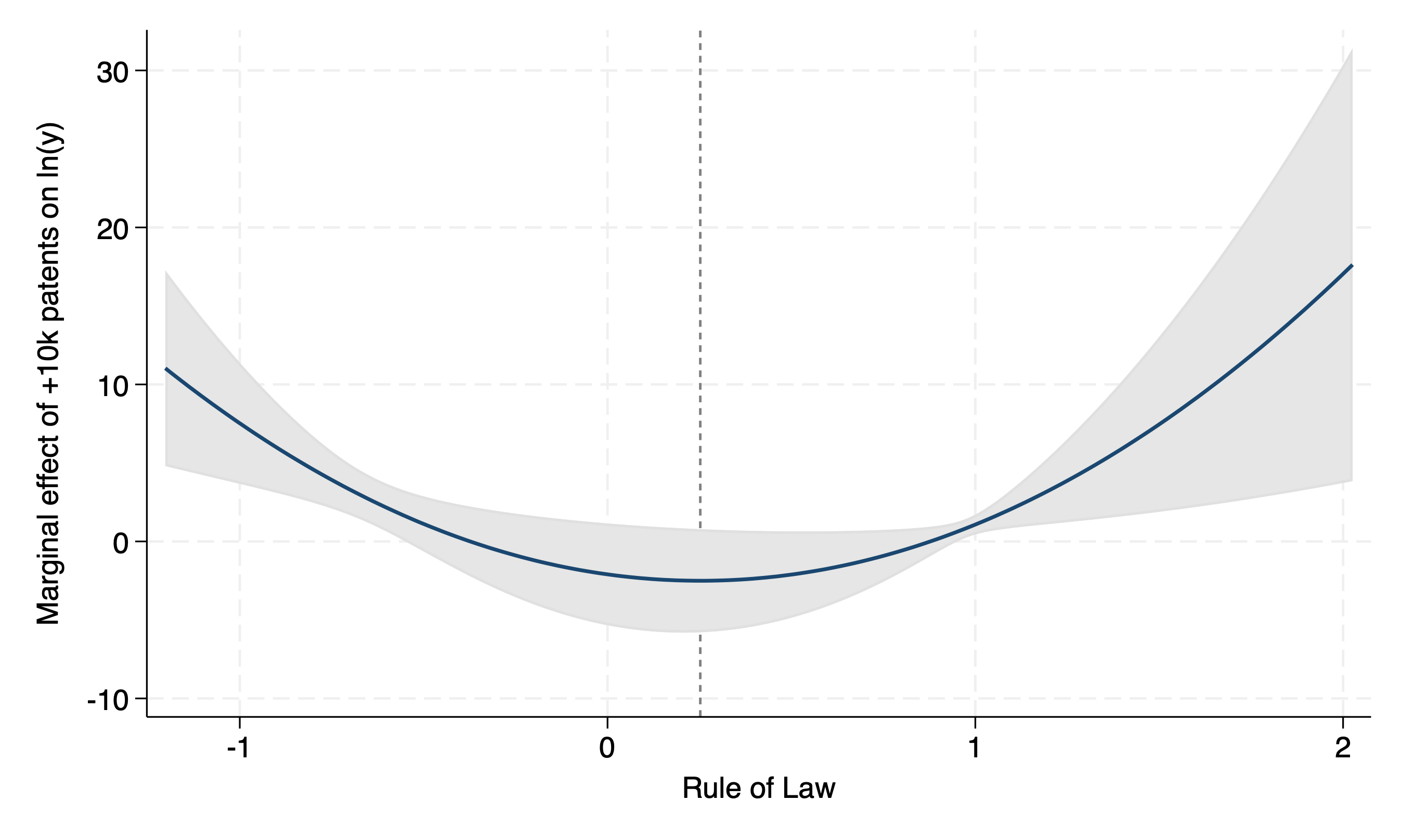}
		\caption{Rule of law}
		\label{fig:efflaw}
	\end{subfigure}	
		\begin{subfigure}{0.49\textwidth}
		\includegraphics[width=\textwidth]{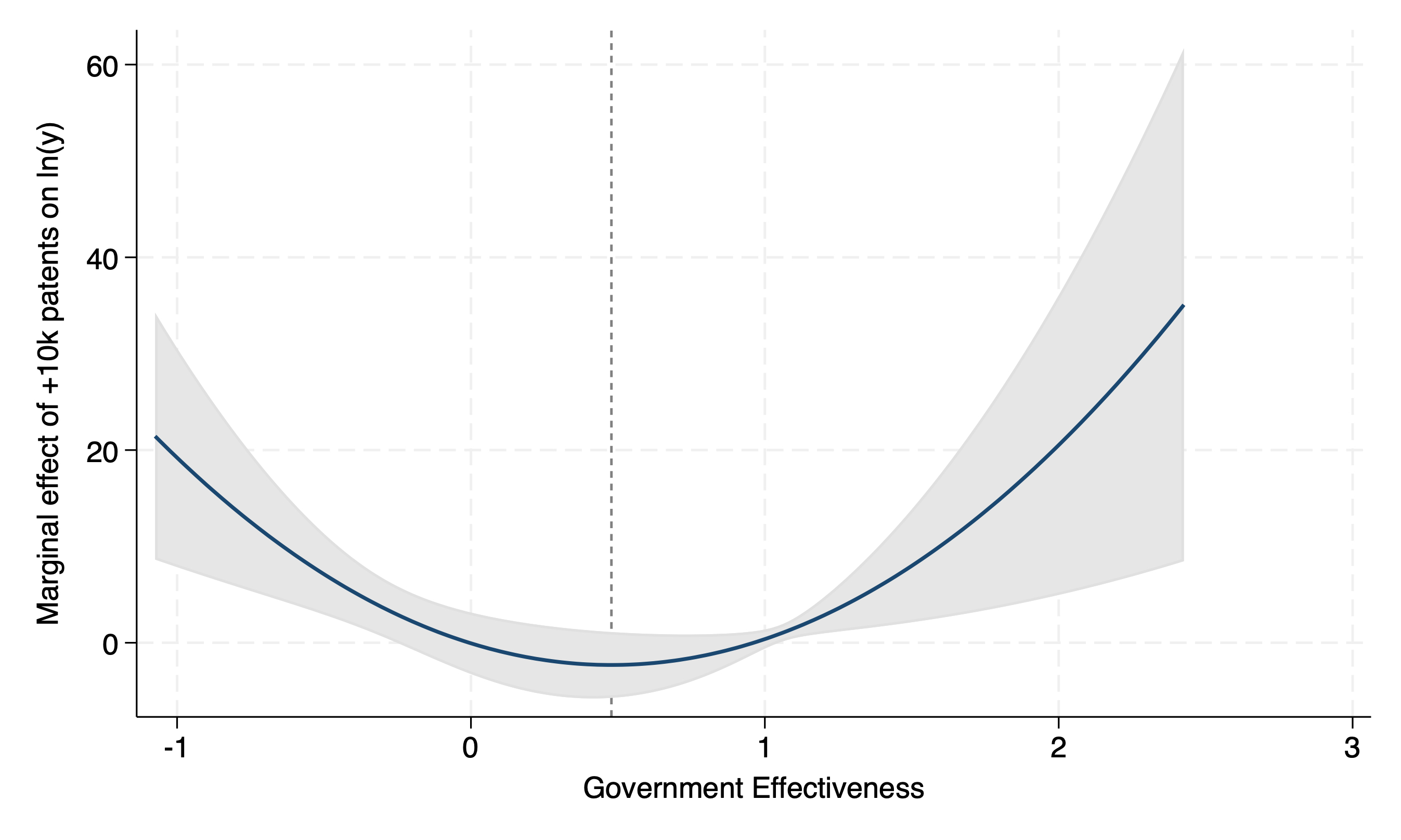}
		\caption{Government effectiveness}
		\label{fig:effgov}
	\end{subfigure}	
	\caption{Nonlinear moderating effect of institutional quality for stock market {\em efficiency} relative to the banking sector excluding China and India}
	\label{fig:ushape}
\end{figure}

Figure \ref{fig:ushape} reports the marginal effect of innovation on stock market efficiency relative to the banking sector.
The moderating pattern is U-shaped for control of corruption (Panel a), rule of law (Panel b), and government effectiveness (Panel c): the marginal effect stays positive at both low and high levels of institutional quality, with a trough at intermediate values. The low-institution upturn is present even when China and India are excluded, indicating that the alternative-financing channel is not specific to these two economies but is a broader feature of low institutional quality environments, consistent with evidence that trust shapes equity-market participation \citep{guiso2008trusting} and that trust- and relationship-based informal finance substitutes for formal mechanisms where enforcement is weak \citep{xu2024inherited}. At the high-institution end, the upturn reflects the role of contract enforcement and reliable public services in lowering information and enforcement frictions, allowing innovation to translate into trading and liquidity \citep{la1997legal,demirgucc2001financial}. The 95\% confidence band nonetheless widens markedly at both extremes of institutional quality, so the magnitude of the moderating effect away from the center should be interpreted with caution.

\section{Conclusion} \label{sec:conclusion}

This paper revisits the relationship between innovation and financial structure from a new perspective. We conceptualize innovation as a catalyst that gradually shapes financial structure of an economy. Drawing on a broad panel of 75 countries over 40 years, we provide empirical evidence that innovation increases stock market activity, efficiency and size  relative to the banking sector. By employing instrumental variables as well as various nonlinear specifications, we address endogeneity concerns and explore the heterogeneous nature of those relationships.

Our results suggest that while innovation leads to more active, more efficient, and larger stock markets relative to the banking sector, the banking sector actively finances innovation when institutional quality is high. This might happen as banks use patents as collateral for loans, increasing banks' assets relative to stock market capitalization.
Furthermore, we find that  institutional quality moderates the relationship between innovation and stock market efficiency relative to the banking sector in a nonlinear way, strengthening the effect at both low and high levels of institutional quality. This uncovers financing channels that are not captured by the law-finance nexus literature.

\appendix
\renewcommand{\thetable}{A\arabic{table}}
\setcounter{table}{0}

\clearpage

\section{Robustness checks}\label{sec:robustness}

In addition to the main findings, this appendix presents several robustness checks, demonstrating that the estimated relationship between innovation and stock market activity, efficiency and size  relative to the banking sector remains stable across key alternative specifications. Further robustness checks, including GMM estimations to probe the sensitivity of the results to different empirical specifications, samples, and estimation methods are provided in the Online Appendix.

\subsection{Alternative financial structure measures}

To test the robustness of our findings, we employ alternative measures for the three dimensions of financial structure. For activity, we replace private credit by deposit money banks as a share of GDP with total private credit extended by both deposit money banks and other financial institutions as a share of GDP. For size, we redefine the size of the banking sector as the combined assets of deposit money banks and central banks as a share of GDP, rather than considering deposit money banks alone. For efficiency, we use overhead costs of banks as a share of total assets, providing an alternative perspective to bank net interest margins as a proxy for operational efficiency.

\begin{table}[ht]
	\centering
	\footnotesize
	\begin{threeparttable}
		\begin{tabular}{
				>{\raggedright\arraybackslash}p{3cm} 
				>{\centering\arraybackslash}m{2.6cm} 
				>{\centering\arraybackslash}m{2.5cm}
				>{\centering\arraybackslash}m{2.6cm}
			} 
			\toprule
			& Activity & Efficiency & Size \\
			\cmidrule(lr){2-2} \cmidrule(lr){3-3} \cmidrule(lr){4-4}
			& (1) & (2) & (3) \\
			\midrule
			
			\underline{2nd Stage} &  &  &  \\
			
			Patents & 0.123**  & 0.576***  &0.163***   \\
			& (0.059) &  (0.119)  & (0.060)   \\
			
			\underline{1st Stage} &  &  &  \\
			
			IV & 
			-1.533***  & -0.442*** &-0.305***     \\
			& 
			(0.289) &  (0.135) & (0.098)    \\
			\underline{Weak IV Test} &  &  &  \\ 
			
			CLR  & 
			0.000 & 0.002 & 0.000 \\
			
			AR  & 
			0.000 & 0.002 & 0.000 \\
			
			Wald  & 
			0.036 & 0.000 & 0.007 \\
			
			\midrule
			Period & 1982-2021 & 1991-2021  & 1982-2021  \\
			Time FE & YES & YES & YES \\
			Controls & YES & YES & YES \\
			Obs. & 955 & 698 &1,036   \\[1ex] 
			\bottomrule
		\end{tabular}
		\begin{tablenotes}
			\footnotesize
			\item The dependent variable is one of three alternative measures for financial structure. The endogenous variable is the number of patent applications instrumented with the regional innovation spillover instrument and regional dummies. Control variables are in Table \ref{table:data_description}. Note: *** \( p < 0.01 \), ** \( p < 0.05 \), * \( p < 0.1 \). Numbers in parentheses are standard errors clustered by year. AR and Wald tests follow the procedures in \cite{olea2013robust}. Multiple IVs yield extra CLR statistics; see \cite{pflueger2015robust} for discussions of weak instrument tests in linear IV regressions and \cite{finlay2014weakiv10} for Stata implementations. $P$-values are reported for CLR, AR, and Wald tests.
		\end{tablenotes}
		\caption{Effect of innovation on alternative measures of financial structure}
		\label{table:alterdependent}
	\end{threeparttable}
\end{table}

Table \ref{table:alterdependent} presents the results using these alternative financial structure indicators. The second-stage estimates show that innovation, proxied by patent applications, remains positively and significantly associated with all three dimensions. Specifically, the coefficient for patents is significant and positive for activity (0.123, significant at the 5\% level), for size (0.163, significant at the 1\% level) and for efficiency (0.576, significant at the 1\% level). These results suggest that higher innovation activity increases stock market activity, efficiency and size relative to the banking sector, even under alternative definitions of financial structure. The first-stage coefficients remain significant for all dimensions, indicating the relevance and strength of the instruments. Weak-instrument tests (CLR, AR, and Wald) yield p-values close to zero in most cases, supporting the validity of the instruments employed.

\subsection{Alternative instruments}

By employing alternative measures for the geographic distance between countries, we construct three additional instruments based on the following methods. First, we use the distance between capitals of countries instead of the distances between the largest cities. The other two measures are weighted calculations. To quantify the weighted distances between countries, we follow the methodology of \cite{head2002illusory}. Let $d_{k,\ell}$ denote the distance between agglomeration $k$ in country $i$ and agglomeration $\ell$ in country $j$. The population-weighted average distance between countries $i$ and $j$ is computed as
\begin{equation*}
	d_{i,j} = \left( \sum_{k \in i} \frac{\text{pop}_k}{\text{pop}_i} \sum_{\ell \in j} \frac{\text{pop}_\ell}{\text{pop}_j} \, d_{k,\ell}^{\theta} \right)^{1/\theta}
\end{equation*}
where $\text{pop}_k$ and $\text{pop}_\ell$ denote the populations of agglomerations $k$ and $\ell$, respectively, and $\theta$ determines how sensitive trade flows are to distance.

For the weighted measure in terms of population, we 
produce a simple arithmetic mean of bilateral distances weighted by population shares by setting $\theta = 1$. By contrast,  we produce a harmonic mean of bilateral distances by setting $\theta = -1$. The harmonic mean assigns greater weight to shorter distances and aligns with the negative coefficients typically observed for distance in gravity models of bilateral trade. Consequently, using $\theta = -1$ better captures the economic proximity, reflecting the fact that nearby regions tend to exhibit disproportionately strong trade and economic linkages.\footnote{We also allow the regional innovation leaders to be time-varying, which select the maximum number of patent application holders in each year within the region. The results are reported in the Online Appendix.}

\begin{table}[ht!]
	\centering
	\footnotesize
	\begin{threeparttable}
		\begin{tabular}{
				p{2.1cm} 
				>{\centering\arraybackslash}m{1cm} 
				>{\centering\arraybackslash}m{1cm}
				>{\centering\arraybackslash}m{1cm}
				>{\centering\arraybackslash}m{1cm}
				>{\centering\arraybackslash}m{1cm}
				>{\centering\arraybackslash}m{1cm}
				>{\centering\arraybackslash}m{1cm}
				>{\centering\arraybackslash}m{1cm}
				>{\centering\arraybackslash}m{1cm}
			} 
			\toprule
			& \multicolumn{3}{c}{Capital distance IV} 
			& \multicolumn{3}{c}{Weighted distance IV} 
			& \multicolumn{3}{c}{Economic proximity IV} \\
			\cmidrule(lr){2-4} \cmidrule(lr){5-7} \cmidrule(lr){8-10}
			& (1a) & (1b) & (1c) & (2a) & (2b) & (2c) & (3a) & (3b) & (3c) \\
			\midrule
			\underline{2nd Stage} & & & & & & & & & \\
			Activity & 
			0.133**  &           &           &
			0.133**  &           &           &
			0.133**  &           &           \\
			
			& 
			(0.057) &           &           &
			(0.056) &           &           &
			(0.056) &           &           \\
			
			Efficiency & 
			         & 0.520*** &  &
			          & 0.514*** & &
			           & 0.514*** & \\
			
			 &
			           &  (0.106)& &
			           &  (0.106)& &
			           &  (0.106)& \\

			Size & 
			&& 0.165**            &
			&& 0.164**             &
			&& 0.163**             \\
			
			& 
			&&  (0.064)            &
			&&  (0.064)            &
			&&  (0.063)            \\
			
			\underline{Weak IV Test} & & & & & & & & & \\
			CLR & 
			0.000 & 0.002 & 0.000 & 
			0.000 & 0.002 & 0.000 &
			0.000 & 0.002 & 0.000 \\
			
			AR & 
			0.000 & 0.002 &0.000  & 
			0.000 & 0.002 &0.000  &
			0.000 & 0.002 &0.000   \\
			
			Wald & 
			0.019 & 0.000  &0.010  & 
			0.019 & 0.000  &0.010  &
			0.019 & 0.000  &0.010  \\
			
			\midrule
			Observations & 
			1,123 & 693 &1,114 &  
			1,123 & 693 &1,114 & 
			1,123 & 693 &1,114  \\
			
			Time period & 
			1982-2021 & 1982-2021 & 1991-2021 & 
			1982-2021 & 1982-2021 & 1991-2021 & 
			1982-2021 & 1982-2021 & 1991-2021 \\
			
			Time FE & 
			YES & YES & YES & 
			YES & YES & YES & 
			YES & YES & YES \\
			
			Controls & 
			YES & YES & YES & 
			YES & YES & YES & 
			YES & YES & YES \\[1ex] 
			\bottomrule
		\end{tabular}
		\begin{tablenotes}
			\footnotesize
			\item The dependent variable is one of three dimensions for financial structure. The endogenous variable is the number of patent applications instrumented with the regional innovation spillover instrument and regional dummies. Columns (1a) to (1c) use the capital distance IV. Columns (2a) to (2c) use the weighted distance IV. Columns (3a) to (3c) use the economic proximity IV. Statistical significance: *** \( p < 0.01 \), ** \( p < 0.05 \), * \( p < 0.1 \). Standard errors clustered by year in parentheses. AR and Wald tests follow \cite{olea2013robust}. See \cite{pflueger2015robust} for discussions of weak instrument tests and \cite{finlay2014weakiv10} for Stata implementations. $P$-values are reported for CLR, AR, and Wald tests of weak instruments.
		\end{tablenotes}
		\caption{Alternative instruments}
		\label{table:instruments}
	\end{threeparttable}
\end{table}

Table \ref{table:instruments} presents the results from using alternative instruments for the number of patent applications, examining whether the baseline findings are robust to different definitions of geographic distance. Across all specifications, the estimated coefficients on the patent variable remain positive and statistically significant, confirming our baseline result.

Specifically, for activity, the coefficients remain very similar across all instruments, consistently around 0.133 with significance at the 5\% level, suggesting that the relationship between innovation and stock market activity relative to the banking sector is stable regardless of the distance measure used. For efficiency and size, we also observe significant coefficients under all three instruments specifications, although magnitudes vary somewhat, particularly for efficiency, which shows larger point estimates (around 0.52). Weak instrument diagnostics (CLR, AR, and Wald tests) consistently produce highly significant p-values close to zero, indicating that the instruments are not weak and lend credibility to the IV estimates. 

In general, these results confirm that the key findings of the baseline analysis are robust to alternative constructions of geographic distance, strengthening confidence in the causal interpretation of the relationship between innovation and stock market activity, efficiency and size relative to the banking sector.

\subsection{Placebo test}

Finally, we present a placebo test by regressing our patent applications on three dimensions of financial structure to address the concern of unobserved common causal factors. As presented in Table \ref{table:placebo}, the coefficients on the first and second lag of activity, efficiency and size are insignificant, which confirms our main results on the causal relationship from patents to stock market activity, efficiency and size relative to the banking sector.

\begin{table}[ht]
    \centering
    \footnotesize
    \begin{threeparttable}
    \begin{tabular}{
        >{\raggedright\arraybackslash}p{4.5cm}
        *{3}{>{\centering\arraybackslash}p{2cm}}
 		}
        \toprule
        & \multicolumn{3}{c}{Patents} \\
        \cmidrule(lr){2-4}
        & (1) & (2) & (3) \\
        \midrule
        Lagged dependent variable & 1.145*** (0.007) & 1.127*** (0.079) & 1.254*** (0.040) \\
        Activity $(t-1)$ & 0.034\quad(0.206) & & \\
        Activity $(t-2)$ & 0.068 \quad(0.205) & & \\
        Size $(t-1)$ & & 0.050 \quad(0.123) & \\
        Size $(t-2)$ & & -0.050 \quad(0.136) & \\
        Efficiency $(t-1)$ & & & 0.171 \quad(0.210) \\
        Efficiency $(t-2)$ & & & -0.122 \quad(0.255) \\
        Financial openness & 0.013 \quad(0.096) & -0.070 \quad(0.110) & 0.029 \quad(0.120) \\
        GDP growth & 0.561 \quad(0.356) & 0.083 \quad(0.141) & 0.050 \quad(0.332) \\
        GDP per capita & -0.377 \quad(0.438) & 0.033 \quad(0.163) & -0.090 \quad(0.333) \\
        Trade openness & -0.223 \quad(0.346) & -0.041 \quad(0.249) & 0.119 \quad(0.285) \\
        Inflation & 0.002 \quad(0.940) & -0.371 \quad(0.968) & -0.798 \quad(3.081) \\
        Government spending & 1.553 \quad(1.276) & 0.130 \quad(0.249) & 0.473 \quad(0.701) \\
        Human capital & -0.076 \quad(0.864) & 0.018 \quad(0.333) & -0.553 \quad(0.627) \\
        Bank Crisis & 0.071 \quad(0.181) & -0.061 \quad(0.278) & -0.240 \quad(0.448) \\
        \midrule
        Observations & 1,168 & 1,165 & 645 \\
        Number of countries & 73 & 70 & 65 \\
        Time-fixed effect & YES & YES & YES \\
        [1ex]
        \bottomrule
    \end{tabular}
    \begin{tablenotes}
         \footnotesize
        \item Note: The dependent variable is the number of patent applications. Statistical significance levels are indicated by the asterisks: *** \( p < 0.01 \), ** \( p < 0.05 \), * \( p < 0.1 \). All estimations are based on the system GMM estimator. Robust standard errors are in parentheses.
    \end{tablenotes}
    \caption{Placebo test}
    \label{table:placebo}
    \end{threeparttable}
\end{table}

\clearpage
\bibliographystyle{apalike} 
\bibliography{ref}

\end{document}